\documentclass[preprint2]{aastex}

\shorttitle{Diffuse Galactic Light from a High Latitude Cloud}
\shortauthors{Ienaka et al.}

\begin{document}


\title{Diffuse Galactic Light in the Field of the Translucent High Galactic Latitude Cloud MBM32 }


\author{N. Ienaka\altaffilmark{1}, K. Kawara\altaffilmark{1},
Y. Matsuoka\altaffilmark{2}, H. Sameshima\altaffilmark{3}, 
S. Oyabu\altaffilmark{2},\\
 T. Tsujimoto\altaffilmark{4},
and B. A. Peterson\altaffilmark{5}}
\affil{
\altaffilmark{1}Institute of Astronomy, University of Tokyo, 2-21-1, Osawa, Mitaka, Tokyo 181-0015, Japan \\
\altaffilmark{2}Graduate School of Science, Nagoya University, Furo-cho, Chikusa-ku, Nagoya 464-8602, Japan\\
\altaffilmark{3}Institute of Space and Astronautical Science, Japan Aerospace Exploration Agency,\\
3-1-1 Yoshinodai, Sagamihara, Kanagawa 229-8501, Japan\\
\altaffilmark{4}National Astronomical Observatory of Japan, 2-21-1 Osawa Mitaka, Tokyo 181-8588, Japan\\
\altaffilmark{5}Research School of Astronomy and Astrophysics, The Australian National University, Weston Creek, ACT 2611, Australia}
\email{ienaka@ioa.s.u-tokyo.ac.jp}

\begin{abstract}
We have conducted {\it B, g, V,} and {\it R}-band imaging in a $45\arcmin \times 40\arcmin $ field containing part of  the high Galactic latitude translucent cloud MBM32, and correlated the intensity of diffuse optical light  $S_{\nu}(\lambda)$ with that of 100$\micron$ emission $S_{\nu}(100\micron)$. A $\chi^2$ minimum analysis is applied to fit a linear function to the measured correlation and derive the slope parameter $b(\lambda )= \Delta S_\nu({\lambda}) / \Delta S_\nu(100\micron)$ of the best-fit linear function. Compiling a sample by combining our $b(\lambda)$ and published ones, we show that the $b(\lambda)$ strength varies from cloud to cloud by a factor of 4.  Finding that  $b(\lambda)$ decreases as $S_{\nu}(100\micron)$ increases in the sample, we suggest that a non-linear correlation including a quadratic term of $S_{\nu}(100\micron)^2$ should be fitted to the measured correlation.  The variation of optical depth, which is $A_V$ = 0.16 - 2.0 in the sample, can change $b(\lambda)$ by a factor of 2 - 3.  There would be some contribution to the large $b(\lambda)$ variation from the forward-scattering characteristic of dust grains which is coupled to the non-isotropic interstellar radiation field   (ISRF). Models of the scattering of diffuse Galactic light (DGL)  underestimate the  $b(\lambda)$ values by a factor of 2. This could be reconciled by deficiency in UV photons in the ISRF or by a moderate increase in dust albedo.  Our $b(\lambda)$ spectrum favors a contribution from  extended red emission (ERE) to the diffuse optical light;  $b(\lambda)$ rises from $B$ to $V$ faster than the models, seems to peak around 6000 \AA\, and decreases towards long wavelengths. Such a characteristic is expected from the models in which the DGL is combined with ERE.

\end{abstract}

\keywords{diffuse radiation -- dust, extinction --- infrared: ISM --- ISM: clouds --- ISM: individual (MBM32) --- scattering}

\section{Introduction}
A diffuse optical component to the Galactic interstellar medium (ISM), which is called diffuse Galactic light (DGL) or sometimes ``optical cirrus,'' was noticed in late 1930's. Extensive studies (e.g., \citealt{elvey1937}; \citealt{henyey1941}; \citealt{vandehulst1969}; \citealt{mattila1979}) revealed that the DGL is starlight scattered off by dust grains in the diffuse ISM which is illuminated by the interstellar radiation field (ISRF).  The spectrum of the DGL extends to ultraviolet (UV) wavelengths where \citet*{witt1997} made extensive studies of the dust scattering properties. The DGL distribution is strongly concentrated toward the Galactic plane. This concentration is attributed to the forward-scattering characteristic of interstellar dust grains coupled with a non-isotropic distribution of stars.

Over 50 years later, diffuse far-infrared (IR) emission, so-called ``IR cirrus,'' was discovered in the {\it IR Astronomical Satellite} ({\it IRAS}) mission \citep{low1984}. It was quickly shown that the IR cirrus is also visible on photographic material (\citealt{devries1985}; \citealt{paley1991}), and suggested that the 100\micron\ brightness varies linearly with extinction in a range of up to $\sim$ 20 MJy sr$^{-1}$ (\citealt{devries1985}; \citealt*{laureijs1987}). Finding good agreement between the structures seen on the optical images and the distribution of the CO emission, \citet{stark1995} pointed out that the optical images can be used to distinguish variations in the dust column density from those in the molecular column density.  These studies indicate that the IR cirrus is the far-IR counterpart of the DGL; far-IR emission in IR cirrus is thermal emission which follows the dust absorption of starlight.  It is thus natural to expect a linear correlation between the DGL and far-IR brightness in the optically thin limit. A combination of optical and far-IR observations would give a powerful tool for investigating the dust properties as well as the ISRF in diffuse ISM (e.g., \citealt{beichman1987} and references therein; \citealt{guhathakurta1989}). 

Since the late 1990s, remarkable progress has been made in the optical and far-IR. In the optical, new wide-field CCD cameras make it possible to cover one or more square degrees at a time, providing the uniform sensitivity over a large area and reducing the systematic uncertainty by temporal airglow variations. In the far-IR, \citet*{schlegel1998}, hereafter SFD98, published the $IRAS/$DIRBE maps, which are reprocessed all-sky maps, by combining $IRAS$ and DIRBE (Diffuse Infrared Background Experiment on board the Cosmic Background Explorer satellite) data.  The $IRAS/$DIRBE maps have DIRBE quality calibration and $IRAS$ resolution and feature removal of point sources as well as of artifacts from the $IRAS$ scan pattern. Recent studies of diffuse optical light benefit from the progress using a wide-field CCD camera and/or the $IRAS/$DIRBE maps. Analysis of diffuse optical light recorded in the $Pioneer$ mission against the $IRAS/$DIRBE 100$\micron$ maps finally led \citet{matsuoka2011} to report the detection of the cosmic optical background (COB). The DGL is an important foreground to be subtracted from data taken to measure the COB. Thus, studying the DGL distribution is indispensable to a better understanding of the COB.   

The present paper mainly focuses on the following issues. The first issue is related to variations in slope parameters, i.e., $\Delta S_\nu(\lambda)/\Delta S_\nu$(100$\micron$), which is determined by fitting a linear function to the measured intensity of diffuse optical light $S_\nu(\lambda)$ as a function of 100$\micron$ intensity $S_\nu$(100$\micron$).  It has been indicated that slope parameters vary from cloud to cloud by a factor of 3 - 4.  For example, \citet{guhathakurta1989} reported the slope parameters scatter by a factor of 3 in their sample of 4 high Galactic latitude clouds, and the slope in a cloud studied by  \citet{paley1991} is 4 times greater than that  obtained from blank-sky spectra sampled in  92,000 directions at high latitude (mostly $|b| >$ 30$^{\circ})$ by \citet{brandt2012}. What does cause such large variations? Another issue is related to extended red emission (ERE) in the diffuse optical light.  ERE detected in the reflection nebula NGC2327 has a broad emission feature with FWHM of 1500\AA\ peaking at 7200\AA\ (\citealt{witt1988}; \citealt{gordon1998}). Carrying out a spectroscopic survey of high-latitude cirrus clouds, \citet{szomoru1998} found that the cirrus spectrum does not agree with that of synthetic scattered E-type galaxy spectra, and explained this mismatch by the presence of ERE. Their cirrus ERE feature peaked at 6000\AA\  and is shifted toward short (bluer) wavelength as compared with ERE in HII regions (peak $\sim$ 8000\AA\ and reflection and planetary nebulae (peak $\sim$ 7000\AA). On the other hand, \citet{brandt2012} concluded that their blank-sky spectrum  is consistent with those expected for their scattering DGL models, finding no evidence of  ERE in the diffuse optical light.  For broadband observations, some authors favor the presence of  ERE in the diffuse optical light (e.g., \citealt{gordon1998}; \citealt{witt2008}; \citealt{matsuoka2011}), while the others suggest the absence of  ERE (\citealt{zagury1999}; \citealt{zagury2006}). 

In the course of our program to constrain the COB, we have carried out broadband imaging of a 45\arcmin \ $\times$ 40\arcmin \ field which contains a translucent molecular cloud at high Galactic latitude. In section 2, we describe the observations and the data processing details, such as flux calibration, masking, and aperture correction. Correlation analysis of the intensity of diffuse optical light against the 100$\micron$ emission is presented in section 3. In section 4, we use our measurements along with similar measurements in the literature to discuss issues affecting the correlation, and consider implications for dust scattering DGL models. A summary appears in section 5.

\section{Observations and Reduction}
\subsection{Observations} \label{sec:observation}
The target field was selected to have the following properties: (1) it is optically thin or translucent at optical wavelengths, and is characterized  by a sufficient contrast at 100$\micron$ to be
distinguished from the 100$\micron$ background;
 (2) it is not crowded with stars and does not include bright stars. The  first criterion is needed in order to see clearly the correlation between optical 100$\micron$ emission. The second is needed in order to minimize the contamination from stars, especially from the extended halos of bright stars. Using these criteria, we have selected the southern part of MBM32, covering $\sim 1\degr \times 1\fdg5$.  This is a diffuse, translucent molecular cloud with  extinction of $A_V= 0.3-0.5$ \citep{magnani1986}, and located at a distance of 120 pc \citep*{magnani1996} at $l$=147\fdg2, $b$=40\fdg7. The surrounding sky is not a pure blank-sky; thermal far-IR emission from interstellar dust was detected. Based on the $IRAS$/DIRBE 100$\micron$ map, \citet{witt2008} estimated $A_V = 0.6$ for the cloud and $A_V = 0.25$ for the surrounding sky.  The 100$\micron$ intensity in the field ranges from  1.4  to 7.0 MJy sr$^{-1}$.

Observations were made at Kiso observatory\footnote{Kiso observatory is operated by the Institute of Astronomy of the University of Tokyo} on dark nights in 2011 February, April, and 2012 February. The Sun and the Moon were more than 35 degrees below the horizon when the data were acquired. The optical data were acquired using the 105cm Schmidt telescope with the  2KCCD camera which has 2048 $\times$ 2048 pixels \citep{itoh2001}. The field of view of the  camera is 50\arcmin \ $\times$ 50\arcmin \ with a pixel scale of 1\farcs5 pixel$^{-1}$. The average seeing was $\sim$ 3\farcs3. The area of high-quality is approximately 45\arcmin \ $\times$ 40\arcmin, smaller than the field of view of the camera due to using dithering.

In Figure 1, the observed  45\arcmin \ $\times$  40\arcmin \ field is indicated with a white square  superimposed on the $IRAS$/DIRBE 100$\micron$ map.  Images were obtained  at four photometric bands, {\it B}, {\it g}, {\it V}, and {\it R}. In order to reduce the effects of  large-scale non-uniformity in the flat-fielding, we observed the field at two different telescope attitudes by flipping the telescope around the right ascension axis in such a way that the difference in hour angle is 180 degrees, thereby rotating the image plane on the detector by 180 degrees. In each telescope attitude, the field was observed three times per  photometric band while dithering a few arcmins. The total exposure time was 1800 s (300 s $\times$ 6) per band. 

\begin{figure}
 \begin{center}
 \includegraphics[scale=0.45]{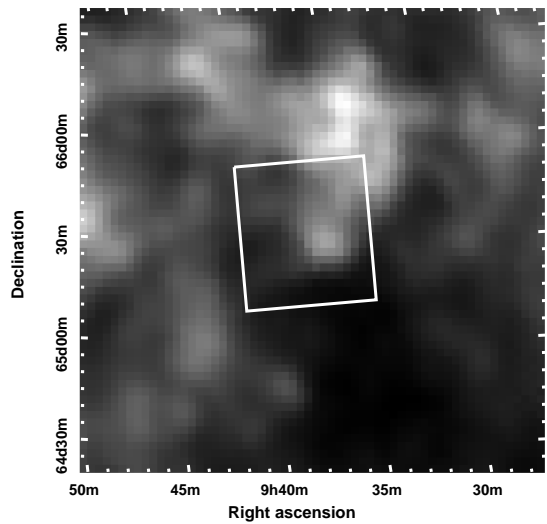}
 \end{center}
 \caption{Optical field superimposed on the $IRAS$/DIRBE 100$\micron$ map. The solid square represents the observed field. The brightest point of the 100$\micron$ map  corresponds to the center of CO emission from molecular cloud MBM32 (\citealt{magnani1985}).}
  \label{fig01}
\end{figure}

\subsection{Data processing}
The data reduction was performed in a standard manner, including overscan subtraction, bias subtraction, and flat-fielding with dome-flat, using the IRAF software package\footnote{IRAF is distributed by the National Optical Astronomy Observatory, which is operated by the Association of Universities for Research in Astronomy (AURA) under cooperative agreement with the National Science Foundation.}.  The flux calibration  was obtained from nearly 1,000 stars in a range of 15 - 19 mag which are listed in the Sloan Digital Sky Survey (SDSS\footnote{Funding for SDSS-III has been provided by the Alfred P. Sloan Foundation, the Participating Institutions, the National Science Foundation, and the U.S. Department of Energy Office of Science. The SDSS-III web site is http://www.sdss3.org/.}) Data Release 8 photometric catalog \citep{aihara2011}.  The SDSS magnitudes ({\it psfMag}) are transferred  to the  {\it BVR} magnitudes  \citep{jester2005}, and then correlated  against instrumental magnitudes to derive the zero-points.  The SExtractor software package \citep{bertin1996}  was used with a circular  aperture of 8$''$ radius.  The uncertainty in the zero-points is estimated to be 0.01 - 0.02 mag. The systematic errors introduced through the conversion of magnitudes are approximately 0.03 mag.

\subsubsection{Masking of stars and galaxies} \label{sec:masking} 
\begin{figure}
 \begin{center}
 \includegraphics[scale=1.0]{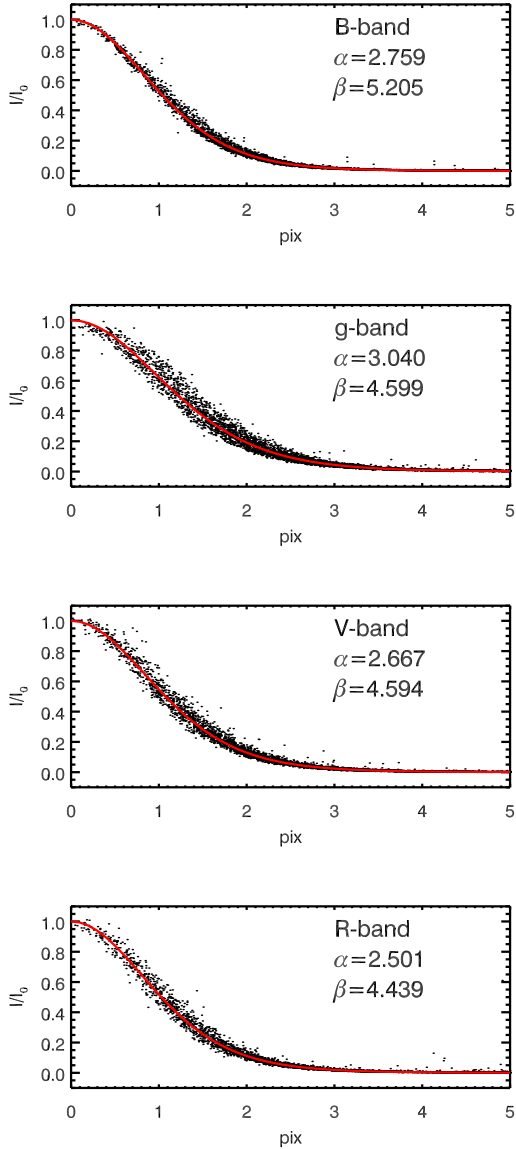}
 \end{center}
 \caption{The scaled radial profiles of stellar images.  Dots represent the normalized fluxes of approximately hundred stars for each band. The red lines are the best fit of  a Moffat function; $I = I_0 [1+(r/\alpha)^{-2}]^{-\beta}$.}
  \label{fig02}
\end{figure}

\begin{figure*}
\hspace{-2em}
 \includegraphics[scale=0.93]{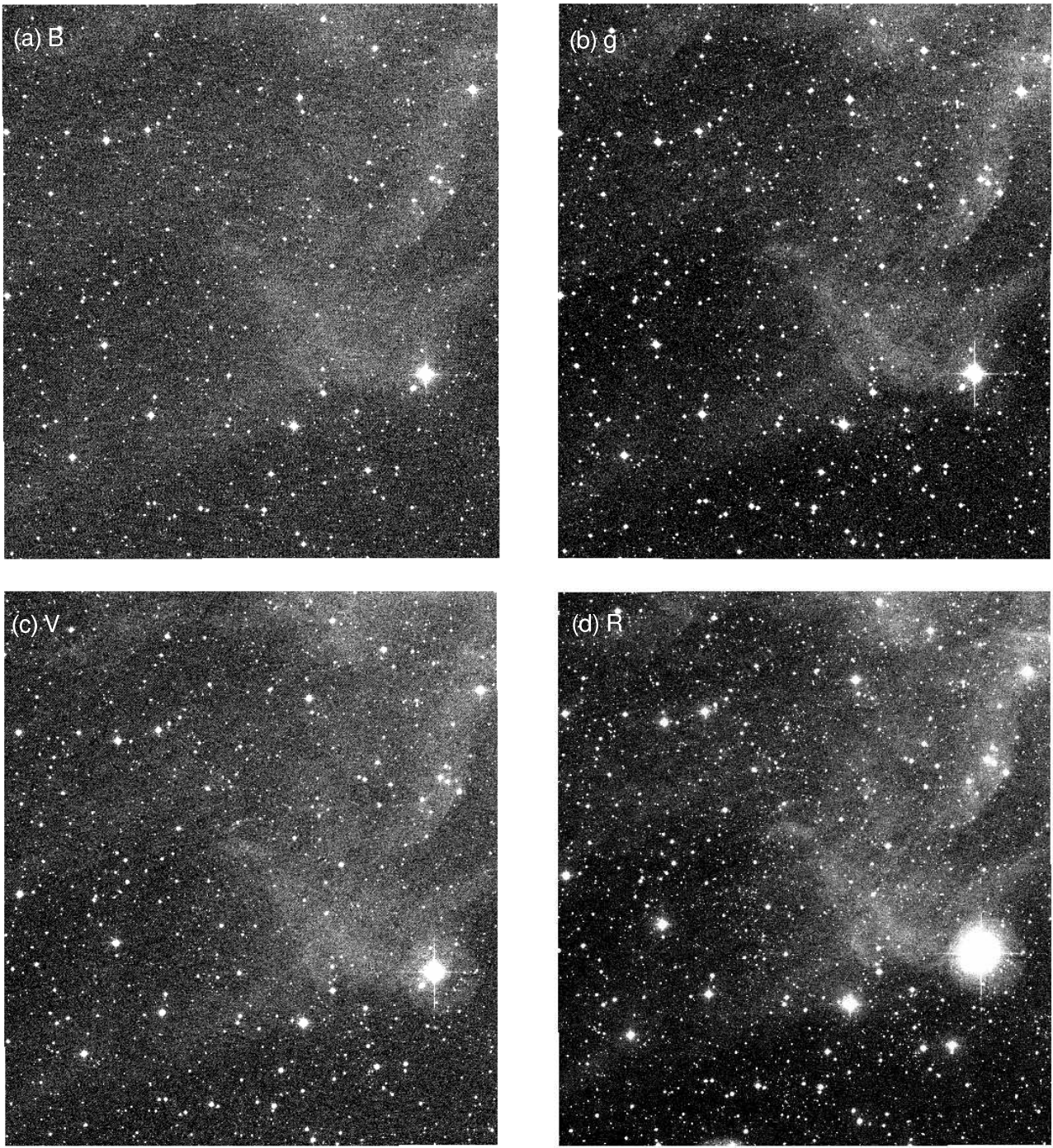}
 \caption{The optical images in {\it B}-band (a), {\it g}-band (b), {\it V}-band (c), and {\it R}-band (d). North is up, east to the left. The field is approximately 45\arcmin $\times$ 40\arcmin, slightly smaller the field of view of instrument due to using dithering.}
  \label{fig03}
\end{figure*}

\begin{figure*}
\hspace{-2em}
 \includegraphics[scale=0.93]{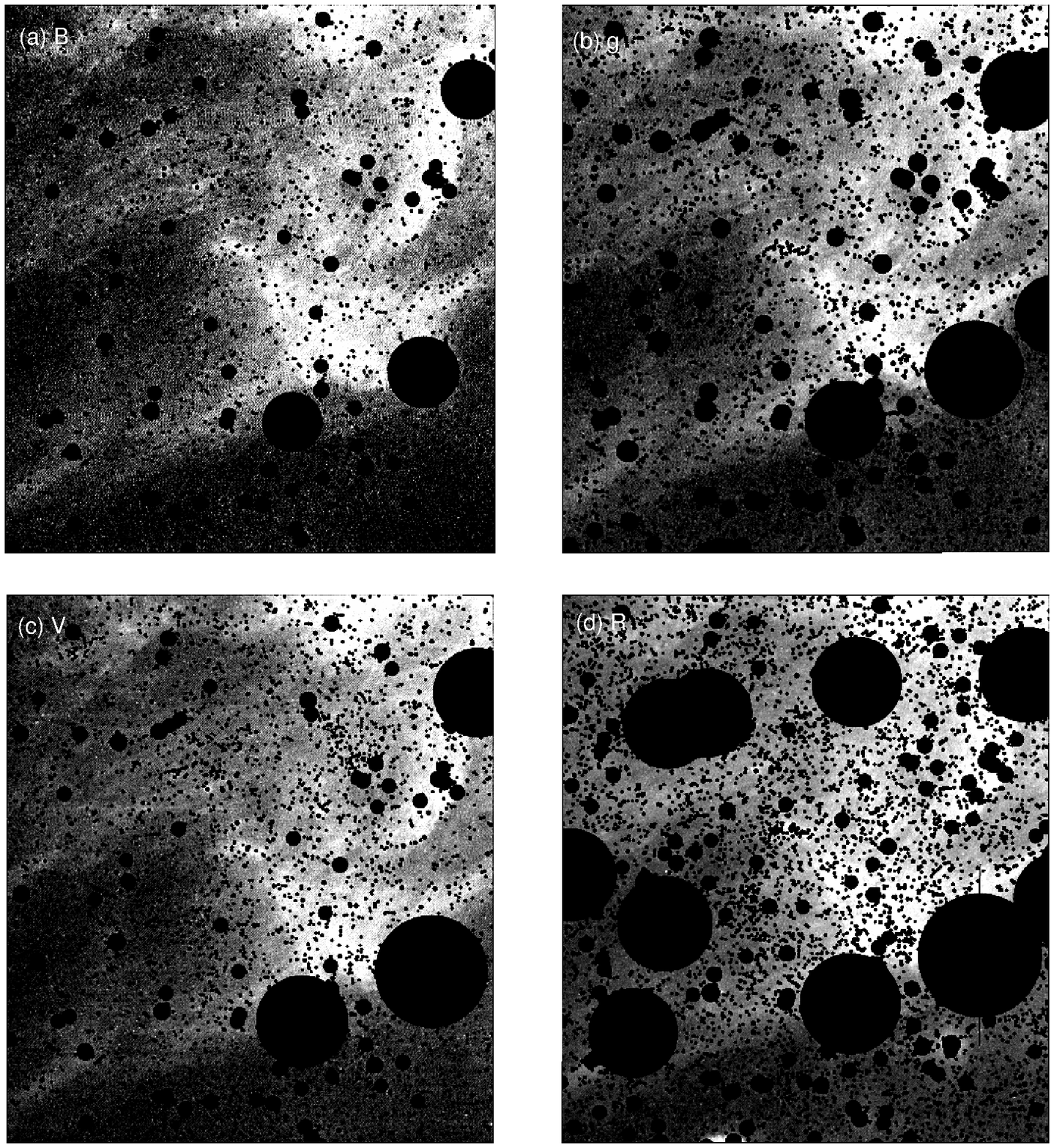}
 \caption{Same as for Figure 3, but for  the masked optical images.}
  \label{fig04}
\end{figure*}

In order to obtain the diffuse component, we masked  stars and galaxies in the  images. Our  masking procedure is as follows. First, we detected all the discrete sources that exceed  1.5 $\sigma$ of the local background  in more than three contiguous pixels using  SExtractor with parameters {\it DETECT\_THRESH} = 1.5 and {\it DETECT\_MINAREA} = 3. Second,  from the detected bright stars, we  derived   point spread functions (PSFs) by fitting a Moffat function $I = I_0[1+(r/\alpha)^{-2}]^{-\beta}$ (\citealt{moffat1969}), where $I_0$ is the intensity at the center of the stellar image and $\alpha$ and $\beta$ are the fitting parameters depending on the seeing.  Figure \ref{fig02} shows the radial profiles of bright unsaturated stars  along with the best-fit curves. Adopting these best-fit curves as the PSF templates, every unsaturated star is masked  with a circle with the radius where the  count level is equal to 1/10 $\sigma$ of the local background count level. Saturated stars,  magnitudes of which are supposed to be underestimated,  are masked with a sufficiently large radius such as five or ten times larger than the masking radius estimated by using the flux value and the PSF for an unsaturated star. The  unmasked and masked images  are presented in Figure \ref{fig03} and Figure \ref{fig04}, respectively. The fractions of masked areas in {\it B, g, V}, and {\it R} images are 17\%, 31\%, 19\%, and 35\%, respectively.

\subsubsection{Aperture correction}
In the flux calibration of the surface brightness of extended sources,  as extensively discussed by \citet{bernstein2002a} and \citet{mattila2003}, one must compensate for the flux that is scattered and diffracted into the aperture. Here we introduce the aperture correction factor, $T_A$, which is the fraction of the flux from a point source within the aperture.  The flux from the point source, $F(\lambda)$, in units of Jy is given by
\begin{equation}
 F(\lambda) = {Q(\lambda) \over T_A(\lambda)} C(\lambda),
\end{equation} 
where $C(\lambda)$ is the signal counts within the aperture in instrumental units (ADU), $Q(\lambda)$ the sensitivity function in units of Jy ADU$^{-1}$. Because $F(\lambda)$ is known and $C(\lambda)$ is measured, we  can only derive $Q'(\lambda) $= $Q(\lambda)$ /$T_A(\lambda)$, but not $Q(\lambda)$ and $T_A(\lambda)$ independently.  In case of an extended source, no flux is lost; the flux lost from the aperture $1-T_A(\lambda)$ is compensated for by the flux scattered in, from the outside the aperture. Therefore, the brightness of the extended source $S(\lambda)$ in units of Jy sr$^{-1}$
is simply given  as
\begin{equation}
 S(\lambda) = {Q(\lambda) \over \Omega} C(\lambda) = {Q'(\lambda) T_A(\lambda) \over \Omega} C(\lambda),
\end{equation} 
where $\Omega$ is the solid angle of the aperture in units of steradians. In the above equation, $S(\lambda)$ is obtained once $T_A(\lambda)$ is given.

The flux loss calculated from the fitted Moffat function is negligible; $T_A(\lambda) = 0.994$. However, \citet{king1971} pointed out that outside the central part of the profile, which is represented by a Gaussian or Moffat function, there is a more slowly declining halo or aureole. The aureole has a profile of an inverse-square  law that extends out from the central part by a factor of 1000 in angular distance, and contains about 5\% of the star's light. Similar results are presented by several authors (\citealt{mattila2003} and references therein). Even higher aureole energy fractions of $\sim 10\%$ are reported in some cases (\citealt{capaccioli1983}; \citealt*{uson1991}). Here, we adopt the aperture correction factor $T_A$ = 0.95$\pm$0.05 as a compromise.

\begin{table}
\begin{center}
 \caption{Observations and errors. \label{tab01}}
 \begin{tabular}{lccc}
  \tableline\tableline
     & $\lambda_{center}$ &  $\sigma(\lambda)$ \tablenotemark{\dagger} & Obs. date \\
Band & ($\micron$) &  (kJy sr$^{-1}$)  & (UT)\\
  \tableline
  {\it B} & 0.44 &  0.48  & 2011-02-04\\
  {\it g} & 0.49 &  0.39  & 2011-04-03\\
  {\it V} & 0.55 &  0.79  & 2011-02-21\\
  {\it R} & 0.65 &  0.39  & 2012-02-21\\
  \tableline
 $IRAS$/DIRBE &  100   &  0.3 $\times 10^3$          & \\
  \tableline
 \end{tabular}
\tablenotetext{\dagger}{Typical standard deviation for a smoothed $2\farcm372 \times 2\farcm372$ pixel.}
\end{center}
\end{table}

\section{Analysis and Results} \label{sec:result}
\begin{figure*}
 \begin{center}
 \includegraphics[scale=0.86]{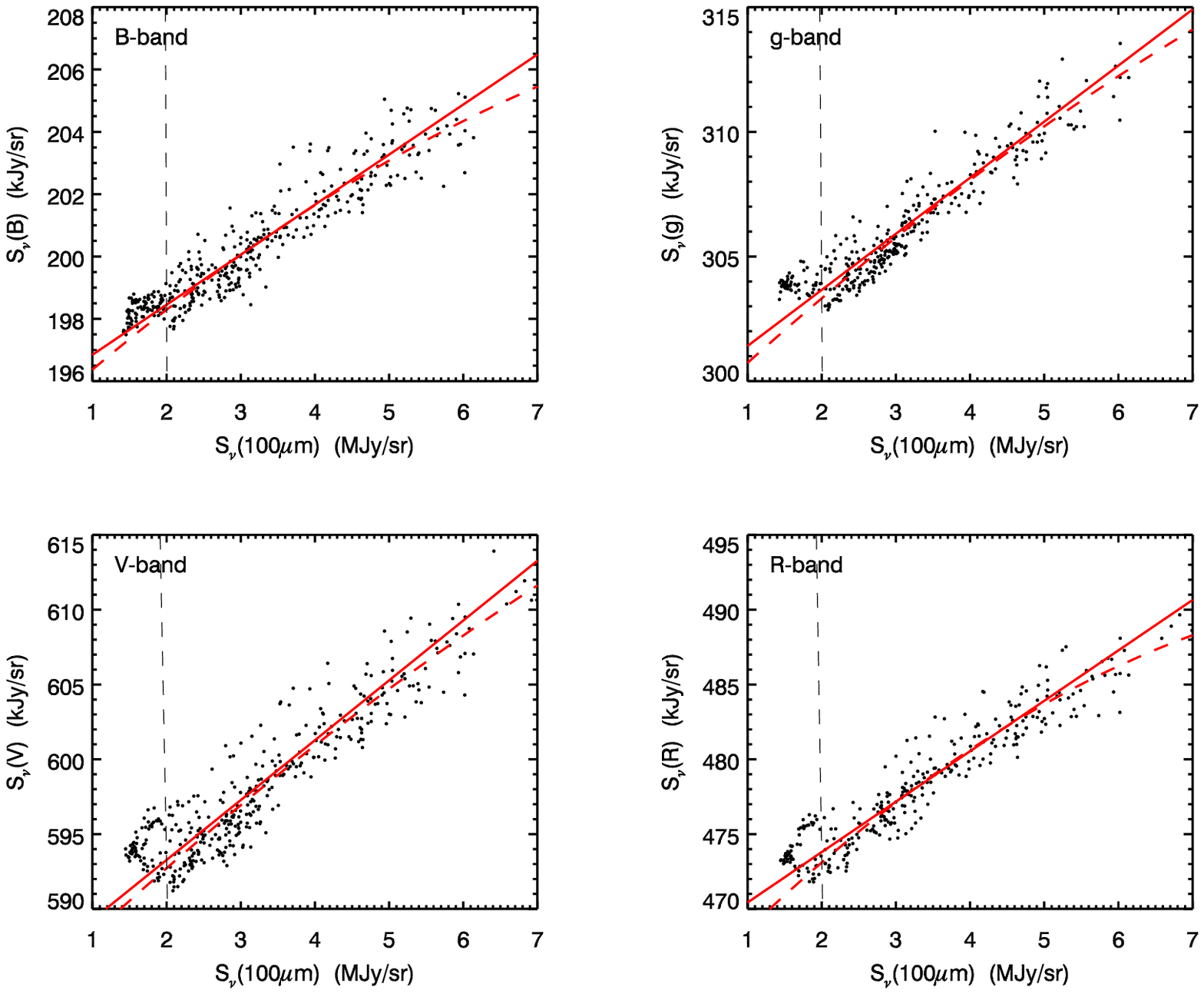}
 \includegraphics[scale=0.86]{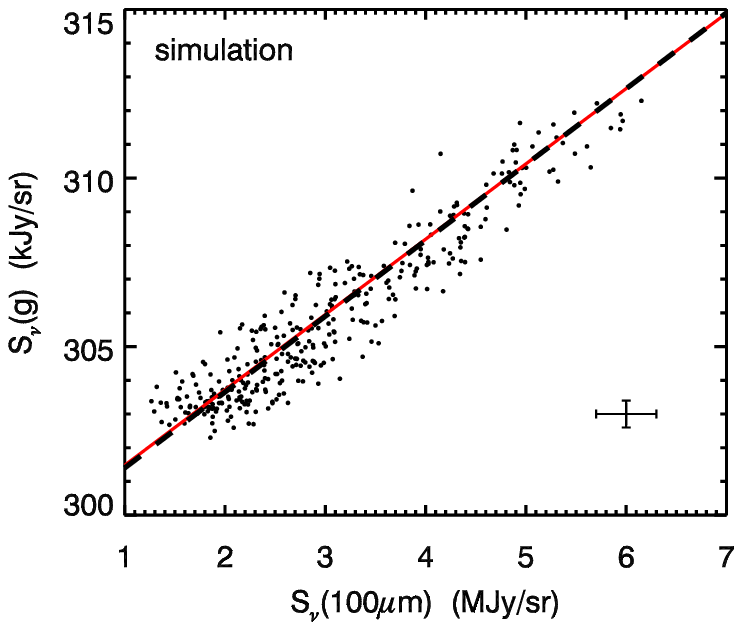}
 \end{center}
 \caption{Correlation of the intensity of the diffuse optical light against 100$\micron$ intensity.  The {\it top} and {\it middle} panels show the measured data. The red solid lines represent the linear functions recovered from the $\chi^2$ minimum analysis, while the red dashed lines represent the quadratic functions.  In case of a quadratic function, the fit is limited to the data points with $S_{\nu}(100\micron$) $\ge$ 2.0 MJy sr$^{-1}$ (right to the vertical dash line); all the data points are used to fit the linear functions. The {\it bottom} panel shows the mock data.  The black dashed line represents the linear function assumed in the simulations, while the red line represents the recovered function. }
  \label{fig05}
\end{figure*}

Figure \ref{fig05}  plots the intensity of diffuse optical light, $S_{\nu}(\lambda)$, as a function of 100$\micron$ intensity, $S_{\nu}(100\micron)$. $S_{\nu}(100\micron)$ is taken from the $IRAS$/DIRBE 100$\micron$ map where the zodiacal light, stars, and galaxies are subtracted. We reduced the resolution of the optical images to the same resolution as the $IRAS$/DIRBE 100$\micron$ map (2\farcm372 $\times $ 2\farcm372) by averaging approximately 9,000 optical pixels. 
Considering the seeing induced correlation, approximately 10 optical pixels constitute one independent data point. Thus,  in the course of smoothing, the optical measurement errors are reduced by a factor of 30 as if averaging 900 independent data points. 
 The mean one $\sigma$ error per smoothed pixel, $\sigma(\lambda)$,
of the smoothed star-masked images is summarized in Table \ref{tab01}.  Finally, we obtained 385, 336, 384, and 309 independent data points at $B, g, V$, and $R$, respectively. 
 
Figure \ref{fig05} shows that $S_{\nu}(\lambda)$ clearly correlates with $S_{\nu}(100\micron$) at all the bands. We now fit to the data a linear function defined as:
\begin{equation}
 S_\nu(\lambda) = a(\lambda) + b(\lambda) S_\nu({100\micron})   \label{eq:linear}
\end{equation}
where the slope parameter at wavelength $\lambda$ is $b(\lambda)$  =  $\Delta S_\nu({\lambda})/\Delta S_\nu(100\micron)$, and the constant parameter $a(\lambda)$ represents  components  independent of  the Galactic diffuse optical light (i.e., atmospheric airglow, zodiacal light, and any other light  such as the COB).  Note that $\lambda$ is $B, g, V, $ or $R$ for our broadband system, and $a(\lambda), b(\lambda) > 0$. 

To perform a minimum $\chi^2$ analysis with the effective variance method \citep{orear1982} which can take into account both x and y errors, we need to know the standard deviation, $\sigma(100\micron$), which is not explicitly estimated by SFD98.
We thus carried out simulations by  generating noiseless data points which follow equation (\ref{eq:linear}) with given $a(\lambda)$ and $b(\lambda)$, and adding Gaussian random errors to these points.  The noiseless 100$\micron$ data range from MIN(100$\micron$) to 8 MJy sr$^{-1}$ with a frequency distribution $\propto S_\nu(100\micron)^{-n}$. At the low intensity end of $S_\nu(\lambda)$ in Figure \ref{fig05}, the correlation appears to be truncated suddenly because of the large scatter of the 100$\micron$ data. The center of the truncated boundary along the $S_\nu(100\micron$) axis is located around 1.8 MJy sr$^{-1}$.  So, we set MIN(100$\micron$) = 1.8 MJy sr$^{-1}$. This was confirmed and justified by changing MIN(100$\micron$) in the simulations.  While $\sigma(\lambda)$  is taken from Table \ref{tab01},  $\sigma(100\micron$)  is a parameter to be determined in the simulations.  We find that a combination of $n = 2$ and $\sigma(100\micron$) = 0.3 - 0.4 MJy sr$^{-1}$ reasonably reproduces the correlation and the frequency distribution as a function of $S_\nu(100\micron$).  

The bottom panel of Figure \ref{fig05} shows the simulated data at the $g$-band along with the linear function recovered from the $\chi^2$ minimum analysis as well as the assumed linear function to generate the mock data points. The top and middle panels plot the observed data along with the recovered linear function.  $a(\lambda)$ and  $b(\lambda)$ are listed in Table \ref{tab02}.

\begin{table*}
\small
\begin{center}
 \caption{DGL model parameters. \label{tab02}}
 \begin{tabular}{lcccccc}
  \tableline\tableline
    & \multicolumn{2}{c}{$S_\nu(\lambda) = a(\lambda)+b(\lambda) S_\nu({100\micron})$} && \multicolumn{3}{c}{$S_\nu(\lambda) =a^Q(\lambda) +  b^Q(\lambda) S_\nu({100\micron}) + c^Q(\lambda)S_\nu({100\micron})^2$}\\
\cline{2-3} \cline{5-7} 
   & $a(\lambda)$ & $b(\lambda)$ && $a^Q(\lambda)$ & $b^Q(\lambda)$& $c^Q(\lambda)$\\
Band & (kJy sr$^{-1}$) & ($\times 10^{-3}$) &&  (kJy sr$^{-1}$) & ($\times 10^{-3}$) & ([kJy sr$^{-1}$]$^{-1}\times 10^{-6}$)\\
  \tableline
  {\it B} &195.23 $\pm$ 0.12 & 1.61 $\pm$ 0.11 && 194.28 $\pm$ 0.50 & 2.17 $\pm$ 0.28 & -0.08 $\pm$ 0.04 \\
  {\it g} &299.16 $\pm$ 0.16 & 2.25 $\pm$ 0.14 && 298.00 $\pm$ 0.60 & 2.80 $\pm$ 0.34 & -0.07 $\pm$ 0.04 \\
  {\it V} &585.27 $\pm$ 0.46 & 4.00 $\pm$ 0.28 && 583.67 $\pm$ 1.02 & 4.74 $\pm$ 0.56 & -0.11 $\pm$ 0.07 \\
  {\it R} &467.06 $\pm$ 0.22 & 3.37 $\pm$ 0.21 && 463.61 $\pm$ 0.91 & 5.23 $\pm$ 0.47 & -0.24 $\pm$ 0.06 \\
  \tableline
 \end{tabular}
\end{center}
\end{table*}

\section{Discussion}
\subsection{Variations in slope parameter $b(\lambda)$}
\begin{figure*}
 \begin{center}
 \includegraphics[scale=0.9]{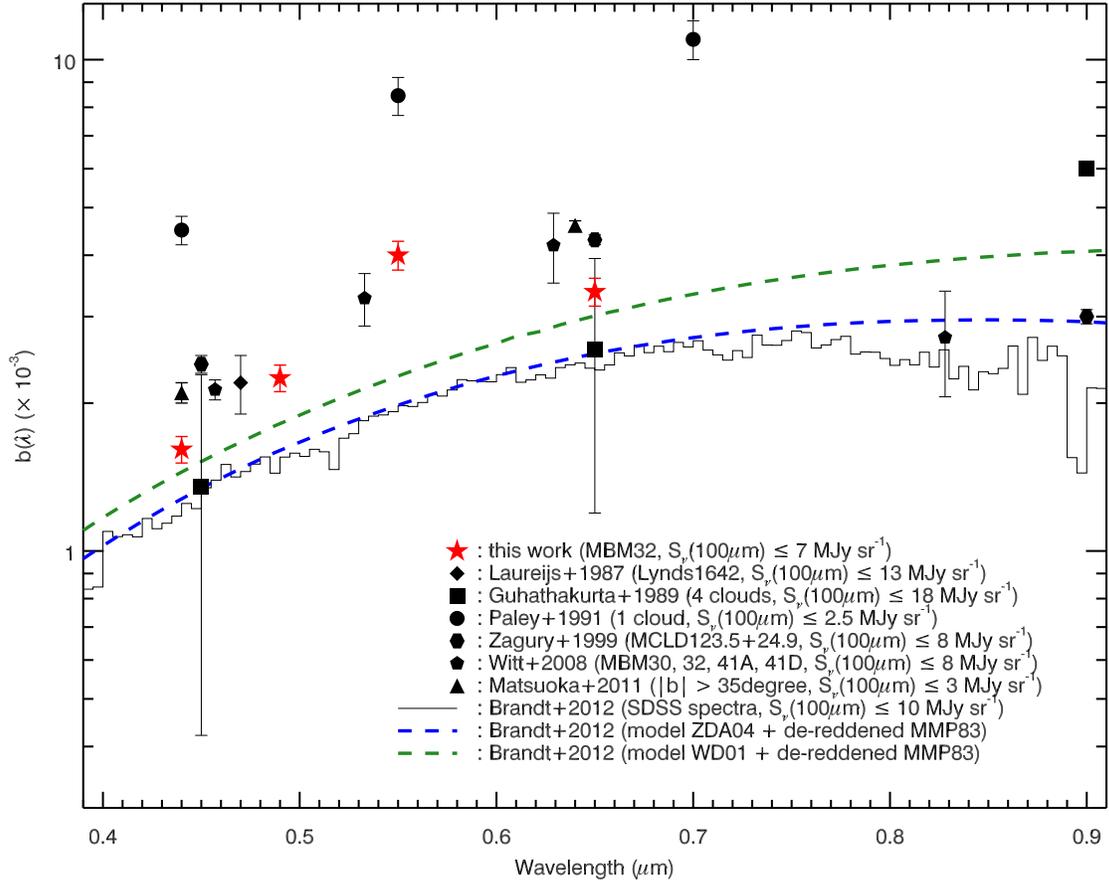}
 \end{center}
 \caption{The correlation slopes $b(\lambda)$  = $\Delta S_{\nu}(\lambda)/\Delta S_{\nu}(100 \micron)$  as a function of wavelengths. }
  \label{fig06}
\end{figure*}

To explore the variations in slope parameter $b(\lambda)$, we compiled $b(\lambda)$ values from the literature into a sample which is given in Table \ref{tab03}. Figure \ref{fig06} compares the data in this sample with model predictions.  For the data points in the sample,  the $IRAS$/DIRBE 100 $\micron$ maps are used in \citet{witt2008}, \citet{matsuoka2011},  \citet{brandt2012}, and this work,  while early work, namely \citet{laureijs1987}, \citet{guhathakurta1989}, \citet{paley1991}, and \citet{zagury1999} used  the original {\it IRAS} 100$\micron$ maps.  The optical data are a collection of heterogeneous samples which were taken with different techniques  and analyzed using  different  methods. The broadband  data taken from the literature are: fields towards diffuse cloud Lynds 1642 studied by \citet{laureijs1987}, four  diffuse clouds ($|b| \ge 25$\degr)  at  {\it $B_J$, R}, and  {\it I}  by \citet{guhathakurta1989}, a compact high-latitude source at {\it B, V}, and {\it R} by  \citet{paley1991}, a molecular cloud MCLD123.5+24.9 by \citet{zagury1999}, and four  high-latitude molecular clouds (MBM 30, 32, 41A, and 41D)  at   {\it B, G, R,} and {\it I} by \citet{witt2008}.  The data studied by \citet{matsuoka2011}  were taken beyond the zodiacal dust zone with  the {\it Pioneer 10/11} Imaging Photopolarimeter(IPP)  at $B$ and $R$ in regions with $S_\nu(100\micron$) $\le$ 3  MJy sr$^{-1}$ at Galactic latitude $|b| > 35\degr$, covering about a quarter of the whole sky. The spectrum  by \citet{brandt2012}, labeled "Full Sky Continuum" in their Figure 3, are obtained by analyzing the optical spectra of nearly 90,000 blank-sky spectra from the Sloan Digital Sky Survey (SDSS), the weighted regional coverage of which is concentrated at intermediate Galactic latitudes $|b| = 30\degr \sim 45\degr$. Their spectrum has a large uncertainty in the flux calibration due to  their analyzing method. The spectra plotted in Figure \ref{fig06}  are scaled with their preferred bias factor of 2.1.  In the case where more than one cloud is published in a single paper, the average $b(\lambda)$ value is plotted with the standard deviation. As can be seen, $b(\lambda)$ varies by a factor of 3 - 4.

In the following subsections, we explore possible causes of the large $b(\lambda)$ scatter  by examining the effects of optical depth, dust albedo, dust temperature and the forward-scattering characteristic of dust grains coupled to the non-isotropic interstellar radiation field   (ISRF).

\begin{table*}
\small
\begin{center}
 \caption{Correlation slopes $b(\lambda)$ for high-latitude clouds or fields. \label{tab03}}
 \begin{tabular}{lcccc}
  \tableline\tableline
Reference &$b(\lambda)^{\dagger}$ & $\lambda_{center}$&$S_\nu(100\micron)$ range\tablenotemark{\ddagger} & $b(R)/b(B)$\\
Clouds or fields          &($\times 10^{-3}$)&(\micron)&(MJy sr$^{-1}$)& \\
\tableline
This work                & & & &  \\
 & & & &  \\
MBM32($l=147, b=41$)               & 1.6 $\pm$ 0.1 & 0.44 & 1 - 7 & 2.1 \\
                        & 2.3 $\pm$ 0.1 & 0.49 & &  \\
                         & 4.0 $\pm$ 0.3 & 0.55 & &  \\
                         & 3.4 $\pm$ 0.2 & 0.65 & &  \\
\tableline
Laureijs et al. 1987     & & & & \\
 & & & &  \\
Lynds1642($l=211, b=-38$)                & 2.2 $\pm$ 0.3 & 0.47 & 1 - 13 & \\
\tableline
Guhathakurta et al. 1989 & & & &  \\
 & & & &  \\
ir1($l=174, b=-42$)                     & 0.4           & 0.45 & 7 - 18 & 3.2 \\
                         & 1.2           & 0.65 & & \\
 & & & &  \\                        
ir2($l=235, b=37$)                    & 1.1           & 0.45 & 3 - 7 & 2.0  \\
                         & 2.2           & 0.65 & & \\
& & & &  \\
ir3($l=38, b=45$)                     & 2.6           & 0.45 & 3 - 9 & 1.7  \\
                         & 4.4           & 0.65 & &  \\
                         & 6.0           & 0.90 & &   \\
\tableline
Paley et al. 1991        & & & &  \\
 & & & &  \\
CTI cirrus cloud($l=104, b=-32$)                 & 4.5 $\pm$ 0.3 & 0.44 & 2.5 & 2.3 \\
                         & 8.5 $\pm$ 0.8 & 0.55 & &  \\
                         &11.0 $\pm$ 1.0 & 0.70 & &  \\
\tableline             
Zagury et al. 1999       & & & &  \\
 & & & &  \\
MCLD123.5+24.9($l=124, b=25$)           & 2.4 $\pm$ 0.1 & 0.45 & 5 - 8   & 1.8 \\
                         & 4.3 $\pm$ 0.1 & 0.65 & &  \\
                         & 3.0 $\pm$ 0.1 & 0.90 & &  \\
\tableline
Witt et al. 2008         & & & &  \\
 & & & &  \\
Derived from 4 clouds\tablenotemark{*}($90 \le l \le 147, 38 \le b \le 41$)              & 2.1 $\pm$ 0.1 & 0.46 & 1 - 8   & 2.0 \\
                         & 3.3 $\pm$ 0.4 & 0.53 & &  \\
                         & 4.2 $\pm$ 0.7 & 0.63 & &  \\
                         & 2.7 $\pm$ 0.7 & 0.83 & &  \\
\tableline
Matsuoka et al. 2011     & & & &  \\
   & & & &  \\
1/4 of the entire sky ($|b| > 35^{\circ}$)         & 2.1 $\pm$ 0.1 & 0.44 & 1 - 3   & 2.2 \\
                         & 4.6 $\pm$ 0.1 & 0.64 & &  \\
\tableline
Brandt et al. 2012       & & & &  \\
   & & & &  \\
SDSS spectra ($30 \le b  \le 45$)\tablenotemark{**}   & 1.2 $\pm$ 0.1 & 0.45 & 1 - 10  & 1.9  \\
                         & 2.3 $\pm$ 0.1 & 0.65 & &  \\
  \tableline
 \end{tabular}
\tablenotetext{\dagger}{The correlation slope is defined as a linear function of $S_\nu(\lambda) = a(\lambda)+b(\lambda) S_\nu({100\micron})$}
\tablenotetext{\ddagger}{$S_\nu(100\micron)$ range refers to the range in which $b(\lambda)$ is derived assuming a linear function.}
\tablenotetext{*}{MBM30($l=142.2, b=38.2$), MBM32($l=147.2, b=40.7$), MBM41A($l=90.0, b=39.0$), MBM41D($l=92.3, b=37.5$)}
\tablenotetext{**}{The weighted regional coverage of the analyzed nearly 90,000 blank-sky spectra is concentrated at intermediate Galactic latitudes $|b| = 30\degr \sim 45\degr$.}
\end{center}
\end{table*}

\subsubsection{Optical depth and dust albedo}
The correlation in Figure \ref{fig05} may appear to be linear.  However, as we discuss below,  linear correlations would not be expected in the case where the optically thin limit is not applicable, and significant  changes in the optical depth along the sightline would cause large variations in $b(\lambda)$ from cloud to cloud.

\begin{figure*}[tbp]
\hspace{-3em}   \includegraphics[scale=1.0]{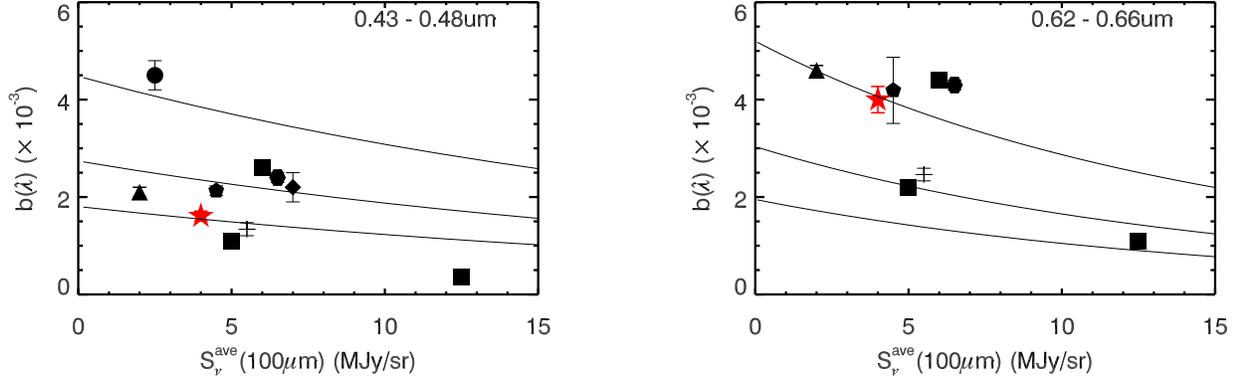} 
   \caption{\small The {\it left}  panel plots the correlation slopes $b(\lambda)$ as a function of $S_{\nu}^{ave}(100\micron$) for the blue range (0.43 to 0.48 \micron). $S_{\nu}^{ave}(100\micron$) refers to the  representative average of the lowest and highest values of the $S_{\nu}(100\micron$) range in which $b(\lambda)$ is derived by fitting a linear function.  Spectroscopic data are averaged over this range.  The {\it right} panel is the same as for the {\it left} panel but for the red range (0.62 to 0.66 \micron). The legend for the points is the
 same as in Figure \ref{fig06}, except for the crosses which are the averaged spectroscopy data from \citet{brandt2012}.  The lines represent the relation expected from equation (\ref{eq:appendix}) for  albedos $\omega_V = 0.6, 0.7,$ and $0.8$ from lower to upper.}
   \label{fig07}
\end{figure*}

Figure \ref{fig07} depicts $b(\lambda)$ as a function of $S_{\nu}^{ave}(100\micron$) in two ranges of the effective broadband wavelengths, namely the blue range(0.43 - 0.48 \micron) and the red range (0.62 - 0.66 \micron). $S_{\nu}^{ave}(100\micron$) refers to the representative average of the lowest and highest values of  the $S_{\nu}(100\micron$) range in which $b(\lambda)$ are derived by assuming a linear function. For the three clouds published in \citet{guhathakurta1989},  the plotted $b(\lambda)$ values refer to individual clouds, not averaged values.  The figure clearly shows that the $b(\lambda)$ decreases as $S_{\nu}^{ave}(100\micron$) increases, indicating that a significant portion of the reason for the large $b(\lambda)$ scatter seen in Figure \ref{fig06} can be attributed to  this $b(\lambda)$ - $S_{\nu}^{ave}(100\micron$)  anti-correlation.

Accordingly we propose a correction to equation (\ref{eq:linear}) as $b(\lambda)$ = $b^0(\lambda) + b^1(\lambda) S_\nu(100\micron)$, or $S_\nu(\lambda)$ = $a(\lambda)$  +  $[b^0(\lambda) + b^1(\lambda) S_\nu(100\micron)] S_\nu({100\micron})$.  Then, $S_{\nu}(\lambda)$ has a quadratic term as follows:
{\small
\begin{equation}
 S_\nu(\lambda) 
               = a^Q(\lambda) +  b^Q(\lambda) S_\nu({100\micron}) + c^Q(\lambda)S_\nu({100\micron})^2, \label{eq:quad}
\end{equation}
}
where $b^Q(\lambda) >0$ and  $c^Q(\lambda)  < 0 $. While the correlation in Figure \ref{fig05} appears to be linear, 
the actual fact is that the correlation is non-linear,
but this is hidden by the prevailing noise. 
Since the quadratic term $ c^Q(\lambda) S_\nu({100\micron})^2$ is negative, this would explain why the correlation of the $Pioneer$ data deviates below the best-fit linear function at the bright end of 100$\micron$ emission \citep{matsuoka2011}. We performed a minimum $\chi^2$ analysis with the effective variance method to fit the quadratic function to our data. As can be seen in Table \ref{tab02} and Figure \ref{fig05}, the derived constant parameter of a quadratic function $a^Q(\lambda)$ is smaller than that of a linear function $a(\lambda)$ at all the photometric bands. Thus, this negative quadratic term would be important for the separation of the COB from the diffuse optical light, reducing the DGL intensity at which $S_{\nu}(100\micron$) is equal to the Cosmic 100$\micron$ background intensity.

\begin{figure}[htbp]
\hspace{-3em}
   \includegraphics{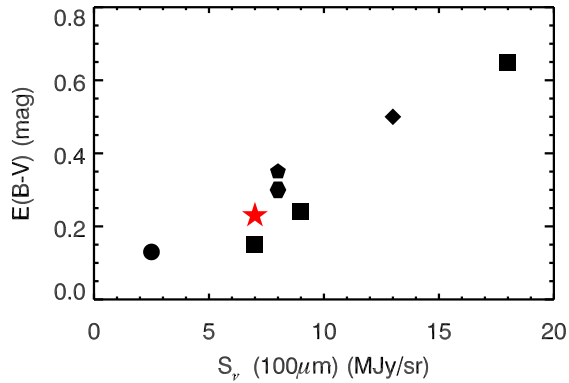} 

   \caption{\small $E(B-V)$ as a function of $S_{\nu}(100\micron$) for the peak of 100\micron\ emission in the cloud. The legend for the points is the same as in Figure \ref{fig06}.}
   \label{fig08}
\end{figure}

The origin of the quadratic term could be attributed to the fact that our sample directions are not optically thin, but translucent or thick.  Figure \ref{fig08} illustrates the relation between reddening $E(B-V)$ and 100$\micron$ emission at the peak of 100\micron\ emission for our clouds, where $E(B-V)$ is taken from the $IRAS$/DIRBE reddening map. If typical extinction curve ($R_V = A_V/E(B-V)= 3.1$) is assumed, the 100$\micron$ brightness per unit of visual extinction is $S_{\nu}(100\micron$)/$A_{V}$ = 8 - 15 MJy sr$^{-1}$ mag$^{-1}$. Our sample ranges from $S_{\nu}(100\micron$) = 3  to 18 MJy sr$^{-1}$,  corresponding  $A_V$ = 0.16 -  2.0 mag.  Thus, dust absorption along the sightline should be taken into account.

To illustrate the effect of extinction along the sightline, as fully discussed in the Appendix, we consider a simple model where a plane-parallel dust slab is illuminated by the monochromatic starlight from the back.  The $b(\lambda)$ is expressed as
{\small
\begin{equation}
b(\lambda) \propto  \exp[-(1-\omega)\tau] [1-\exp(-\omega\tau)] / \{1 - \exp[-(1-\omega)\tau]\}, \label{eq:appendix}
\end{equation}
}
where $\omega$ is the albedo. 

Equation (\ref{eq:appendix}) implies that $b(\lambda)$ depends on the optical depth and the albedo, but not on the ISRF intensity.  In Figure \ref{fig07}, we plot equation (\ref{eq:appendix}) with a single constant value of proportionality\footnote{$b(\lambda)$ = $X \exp[-(1-\omega)\tau] [1-\exp(-\omega\tau)] / \{1 - \exp[-(1-\omega)\tau]\}$}, $X$, both for the left and right panels, assuming $S_{\nu}(100\micron)/A_V =$ 10 MJy sr$^{-1}$ mag$^{-1}$.  $X$ is set to an appropriate value by eye.  The three lines depict the relation for albedo $\omega_V$ of 0.6, 0.7, and 0.8 from bottom to top.  Figure \ref{fig07} suggests that a range of $A_V$ = 0.16 -  2.0 mag could change the $b(\lambda)$ strength by a factor of 2 - 3 while keeping $\omega$ constant.  It also indicates that the $A_V$ range alone cannot explain the full range of the $b(\lambda)$ scatter. If the albedo becomes greater as the optical depth becomes smaller, then the full range of the $b(\lambda)$ scatter could be explained in terms of the optical depth effect coupled to variations in albedo. However, this case is not likely, because small-size dust  grains, which have smaller albedo than large ones(e.g. \citealt{draine2003}), are expected in smaller optical depth regions according to the theory of grain growth.  

Two smooth curves in Figure \ref{fig06} are synthesized spectra for $\tau_{V}$ = 0.15 by \citet{brandt2012}, adopting the ISRF spectrum by \citet{mathis1983}  combined with a simple plane-parallel galaxy model and two dust models from \citet*{zubko2004} and  \citet{weingartner2001} (hereafter ZDA04 and WD01 models). The models underestimate $b(\lambda)$ by a factor of 2. There would be two possible explanations for this. One is a deficiency in UV photons compared to the assumed spectrum of the ISRF; the dust temperature is low and $b({\lambda})$ thus increases where the dust grains are illuminated by a small number of UV photons(see Figures 9 and 10 in \citealt{brandt2012}). The other is that the albedo is greater than assumed in the models, because Figure \ref{fig07} and equation (\ref{eq:appendix}) imply that the  $b(\lambda)$ strength is doubled if $\omega$ increases by 20\%; for example, $\omega_V$ to be 0.8 instead of a standard Milky Way value of 0.67 at V \citep{draine2003}. If this is the case, more large grains are needed than in the ZDA04 and WD01 models, because large grains are efficient scatters of light.

\subsubsection{Dust Temperature}

Another possible cause of the large $b(\lambda)$ scatter is variations in dust temperature.  Higher dust temperature increases the intensity of  the 100\micron\ emission and, as a consequence, it would decrease the $b(\lambda)$. The temperature dependence of the resulting 100\micron\ radiation is $\sim T^{5.8}$ at 18K \citep{brandt2012}. However the real situation is not so simple. The temperature of 100\micron-emitting dust grains, i.e. large grains, is determined by the balance between incoming energy from the illuminating radiation to the grains and the outgoing infrared thermal energy. One can increase the temperature of the grains by increasing the UV component of the ISRF while leaving the optical intensity of the ISRF the same. That would leave the optical DGL intensity the same but increase the 100\micron\ intensity. However, if the optical intensity of the ISRF changes, it also changes the intensity of the DGL. Thus, the increment of $b(\lambda)$, e.g the ratio of increment of the DGL and that of the 100\micron\ emission, depends intricately on the spectral energy distribution and the non-isotropy of the ISRF. \citet{lehtinen2007} showed that the dust emissivity, which is measured from the ratio of far-IR optical depth to visual extinction $\tau$(far-IR)/$A_V$, decreases as a function of dust temperature, and the higher temperature dust does not necessarily mean the higher surface brightness at 100\micron. They concluded that the emissivity variations are caused by the variations in the absorption cross section of dust in the far-IR, or the grain size distribution. 

\begin{figure*}[tbp!]
\hspace{-3em}   \includegraphics[scale=1.0]{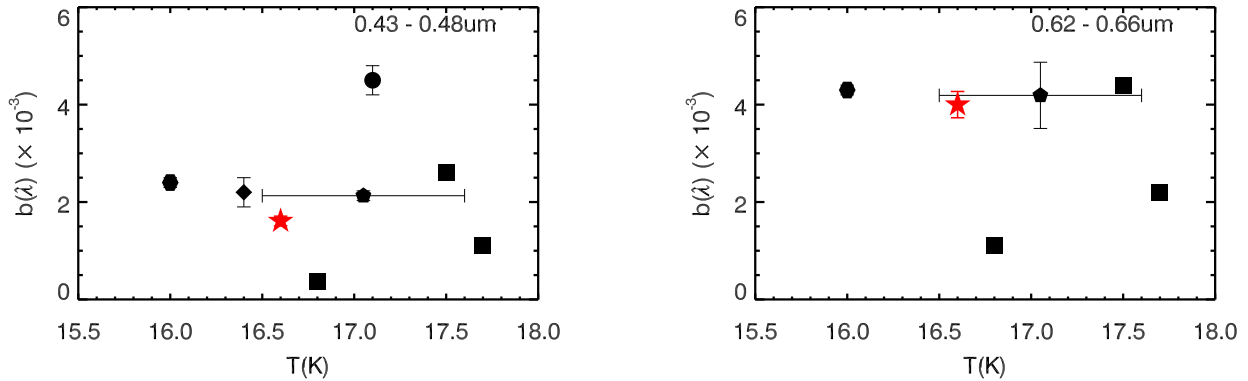} 
   \caption{\small The {\it left}  panel plots the correlation slopes $b(\lambda)$ as a function of $T_{dust}$ for the blue range of the effective broadband wavelengths. The {\it right} panel is the same as for the {\it top} panel but for the red range. The legend
for the points is the
same as in Figure \ref{fig06}. The horizontal error bars are represent their temperature ranges among the clouds.
   \label{fig09}}
\end{figure*}

In our sample, we do not see any convincing correlation between $b(\lambda)$ and dust temperature $T$ as shown in Figure \ref{fig09}. The plotted dust temperatures are taken from the $IRAS$/DIRBE temperature map. The $T^{5.8}$ dependence of the 100\micron\ intensity causes an increase by a factor of about 1.7 over a temperature range from 16K to 17.5K shown in the data of Figure \ref{fig09}. However, the ratio of $\tau$(far-IR)/$A_V$ declines with increasing temperature, and the effective increase in $S_{\nu}(100\micron$) would become lower than 1.7. This may explain why the data of Figure \ref{fig09} do not lead to any discernible trends. As a caveat, it should be noted that the $IRAS$/DIRBE temperature map is severely limited by the angular resolution of DIRBE of $\sim$ 1 degree. Thus, this map is not capable of correctly reflecting the temperature variations across cloud structures with scales much smaller than the DIRBE resolution.  Further exploration of the $b(\lambda)$ dependence on the dust temperature requires far-IR data with higher resolution such as those from the Akari all sky survey. Such analysis is beyond the scope of this paper.

\subsubsection{Forward-scattering characteristic of dust grains}

Another possible cause of the large $b(\lambda)$ scatter is the effect of the forward-scattering characteristic of dust grains. The ISRF is non-isotropic because of the stellar concentration toward the Galactic disk. The scattering efficiency in the direction towards the observer is  coupled to this non-isotropy through the forward-scattering phase function of scattering grains. This effect is maximized for the geometric configuration explored by 
\citet{jura1979} and \citet{stark1995}.  The latitude dependence of surface brightness of a cloud at high latitude was studied, employing the \citet{henyey1941} approximation for the scattering phase function. In this model, it is assumed that a cloud is illuminated by an infinite homogeneous disk. The observer is located in the plane, and the cloud is located above the plane. The numerical results by \citet{jura1979} show that the surface brightness of a cloud $S$ is written as 
\begin{equation}
S \propto 1-1.1g \sqrt{\sin |b|} \label{jura}
\end{equation}
where $g$ is the asymmetry factor of the phase function and $b$ the Galactic latitude. For strong forward scattering, $g \sim 1$; for isotropic scattering, $g \sim 0$. This equation shows that the brightness of the scattered light is significantly changed with the latitude of the cloud. Adopting a typical asymmetry factor $g \sim 0.75$, the intensity of the scattered light at  $b = 30\degr$ is 2-3 times larger than that of at $b = 90\degr$. This model assumes that all radiation sources, i.e. stars, are located just in a plane, however, this is not the case.  In fact, the mean vertical height of high-latitude clouds is $\sim 150$ pc \citep{magnani1996} and that of the stellar disk is $\sim 300$ pc \citep{bahcall1980, gilmore1983}. Thus, the geometry used in the model of \citet{jura1979} is extreme. It would be more natural to assume that a significant portion of stars are located behind the high-latitude clouds.

\begin{figure*}[tbp!]
\hspace{-3em}   \includegraphics[scale=1.0]{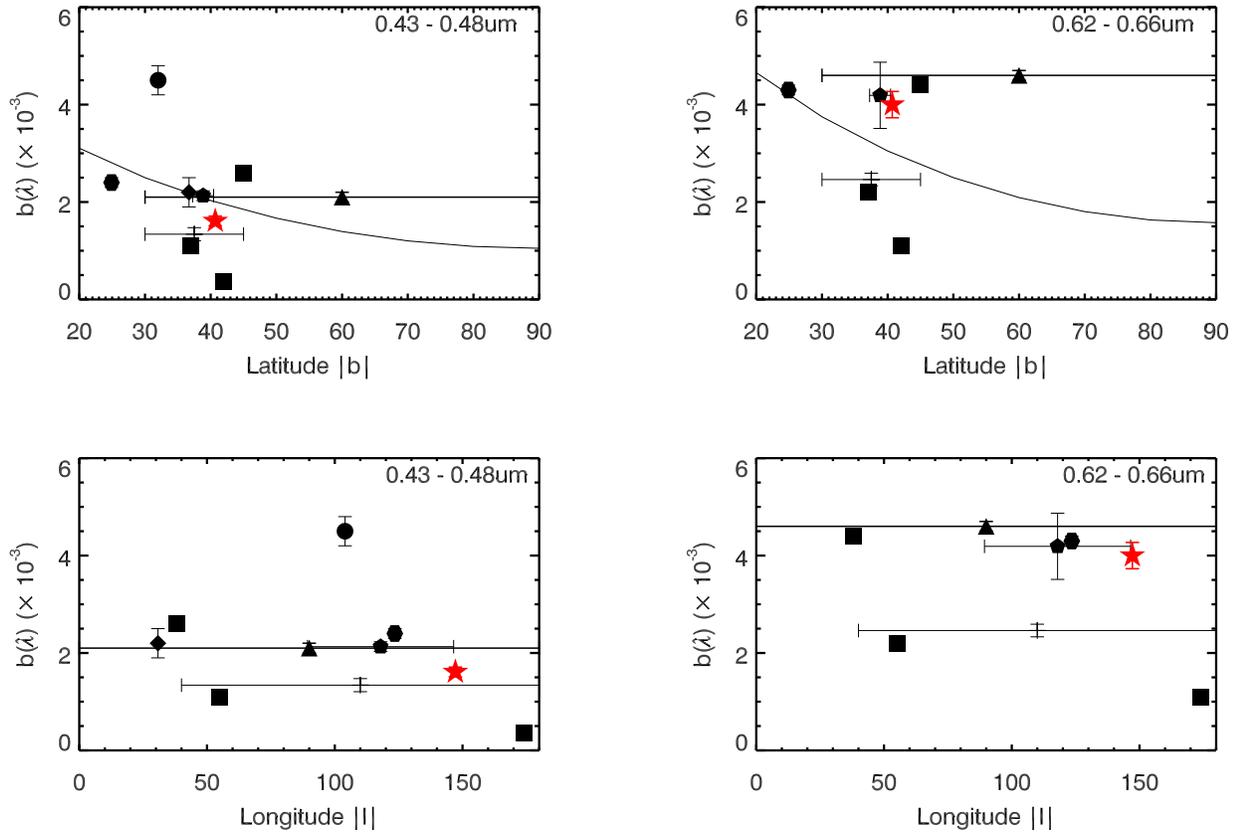} 
   \caption{\small The {\it upper} panels plot the correlation slopes $b(\lambda)$ as a function of Galactic latitude $|b|$, and the {\it lower} panels the correlation slopes $b(\lambda)$ as a function of Galactic longitude $|l|$.
The {\it left} and {\it right} panels are for the blue range of the effective broadband wavelengths and for the red range, respectively. The line solid curve represents the prediction from equation (\ref{jura}), which is derived by assuming  models where the effectiveness of the forward-scattering characteristic of dust grains is maximized. The legend
for the points is the 
same as in Figure \ref{fig06}. The horizontal error bars are represent coordinate ranges.
   \label{fig10}}
\end{figure*}

To examine the effectiveness of the forward-scattering characteristic of dust grains, we present Figure \ref{fig10}, illustrating $b(\lambda)$ as a function of Galactic latitude $|b|$ ({\it upper panels}) and longitude $|l|$ ({\it lower panels}). In upper panels, we  may see a correlation between the $b(\lambda)$ and the Galactic latitude $|b|$ in the blue range, but no correlation in the red range.  There are no $b(\lambda)$ dependency on the longitude $|l|$ at all.  For comparison, equation (\ref{jura}) with $g = 0.75$ is plotted in upper panels of Figure \ref{fig10}. The absolute value of the equation is determined by eye to fit the observed data. The $b(\lambda)$ variation in the blue range seems to be partly explained in terms of the forward-scattering. Then, why are there no correlations in the red range? A possible cause could be a difference in the distribution of blue stars (i.e., effective emitter of blue light) and red stars (i.e., effective red light emitter).  Blue young stars are more concentrated towards the Galactic plane than red old stars and the $g$ value at the blue range is greater than the red range in dust models (e.g, \citealt{draine2003}). Thus, the effect of the forward-directed scattering phase function would be more efficient in the blue range than in the red range. 

\subsection{Extended red emission}

Extended red emission (ERE) is observed as an excess in the 5000-9000\AA\  spectral range in the ISM. It is suggested that ERE is the result of an interaction of far-UV photons with interstellar dust \citep{witt1985, darbon1999, witt2006}. ERE is observed in reflection nebulae \citep{witt1984, witt1988, witt1990}, HII regions \citep{perrin1992}, and planetary nebulae \citep{furton1990,furton1992}. In the diffuse ISM, such as high latitude clouds, some authors favor the presence of ERE in the diffuse optical light\footnote{In this section, we use the term ``DGL''  as the scattered component and ``diffuse optical light'' as the total optical component including the DGL and any other emissions.} (e.g., \citealt{guhathakurta1989}; \citealt{gordon1998}; \citealt{witt2008}; \citealt{matsuoka2011}), while others suggest the absence of ERE (\citealt{zagury1999}; \citealt{zagury2006}) based on broadband observations.  The cloud observed by \citet{zagury1999} and \citet{zagury2006} is illuminated by the nearby star Polaris (F7 Ib). The optical surface brightness of the cloud is dominated by scattered light of Polaris, but this star is not blue enough to emit sufficient far-UV photons for ERE excitation (private communications with the anonymous referee; see section 6.4 of \citet{lehtinen2013} for further discussion). Thus, it would be natural for them to miss ERE. Carrying out spectroscopy of the diffuse optical light toward a high latitude cloud, \citet{szomoru1998} claimed to detect ERE peaking at 6000\AA, which is blue-shifted compared with 8000\AA\ for HII regions and 7000\AA\ for reflection nebulae and planetary nebulae. On the other hand, the more recent spectroscopic study by \citet{brandt2012} cannot find any evidence of ERE. 

There is some confusion in the DGL models, which are sometimes used for the claim of ERE detection; ERE is detected, if the observed diffuse optical light is redder than predicted by the model, and vice verse. Here we introduce the $b(R)/b(B)$ value, which is a measure of the redness of the diffuse optical light. As indicated in Table \ref{tab03}, measured $b(R)/b(B)$ values are typically $\sim 2$. Models by \citet{witt1985}, \citet{guhathakurta1989}, and \citet{bernstein2002a} predict $b(R)/b(B) \sim 1.4$, which is bluer than the measured values  $b(R)/b(B) \sim 2$, supporting the presence of ERE in the diffuse optical light.  In contrast, \citet{gordon1998} predicted $b(R)/b(B) \sim 1.9$, using the Witt-Petersohn DGL model(WP model; \citealt{witt1994}; \citealt{witt1997}), an advanced Monte Carlo multiple scattering model utilizing the actual Galactic radiation field and realistic dust scattering properties. \citet{brandt2012} also predicted $b(R)/b(B) \sim 1.9-2.0$ using their scattering DGL models(ZDA04 and WD01). These are consistent with the observed values without ERE. However, the observed values deduced from $PIONEER$10/11 data from the high-latitude regions by \citet{gordon1998} are $b(R)/b(B) \sim 3.2$, which is higher than other observed values, and still redder than the models.

It is emphasized that $b(\lambda)$ in this work on cloud MBM32 in Figure \ref{fig06} has an excess at the $V$-band and favors ERE in the diffuse optical light;  $b(\lambda)$ rises from B to V faster than the models, seems to peak around 6000 \AA\, and decreases towards long wavelengths. Such a characteristic is expected from the models in which scattered DGL is combined  with  ERE.

\section{Summary}
We have conducted {\it B, g, V,} and {\it R}-band imaging in a $45\arcmin \times 40\arcmin$ field containing part of the high Galactic latitude translucent cloud MBM32, and correlated the intensity of the diffuse optical light  $S_{\nu}(\lambda)$ with that of 100$\micron$ emission $S_{\nu}(100\micron)$.  $S_{\nu}(100\micron$) ranges from 1.4 to 7.0 MJy sr$^{-1}$ in the field. A minimum $\chi^2$ analysis is applied to fit a linear function to the measured correlation and to derive the slope parameter $b(\lambda)$ of the best-fit linear function. 

Compiling a sample by combining our $b(\lambda)$ and published ones, we showed that the $b(\lambda)$ strength varies from cloud to cloud by a factor of 4.  We also found that $b(\lambda)$ decreases as $S_{\nu}(100\micron)$ increases in the sample. To incorporate this  $b(\lambda)$ - $S_{\nu}(100\micron)$ relation, we proposed that a non-linear correlation including a negative quadratic term of $S_{\nu}(100\micron)^2$ be fitted to the measured correlation.  

We explored possible causes of the large $b(\lambda)$ scatter by examining the effects of the optical depth, dust albedo, dust temperature and the forward-scattering characteristic of dust grains coupled to the non-isotropic interstellar radiation field(ISRF).  The variation of optical depth, which is $A_V$ = 0.16 - 2.0 in the sample, can change $b(\lambda)$ by a factor of 2 - 3 while keeping the albedo, $\omega$, constant. This indicates that the $A_V$ range alone cannot explain the full range of the $b(\lambda)$ scatter.  Our finding of  a correlation between the $b(\lambda)$ and Galactic latitude $|b|$ in the blue range (0.43 - 0.48\micron) would imply that there would be some contribution to the $b(\lambda)$ scatter from the effect of the forward-scattering characteristic of dust grains coupled to the non-isotropic interstellar radiation field(ISRF).

The models by \citet{brandt2012} underestimate $b(\lambda)$ by a factor of 2.  There would be two possible explanations for this. One is a deficiency in UV photons compared to the assumed spectrum of the ISRF, because fewer UV photons result in lower dust temperature, which increases $b({\lambda})$. The other is that the albedo is greater than assumed in the models. If this is the case, a greater number of  larger grains are needed in the models, because large grains are more efficient scatterers of light.

 Our $b(\lambda)$ spectrum favors the presence of extended red emission(ERE) in the diffuse optical light;  $b(\lambda)$ rises from B to V faster than the models, seems to peak around 6000 \AA\, and decreases towards longer wavelengths. Such characteristic is expected from the models in which the DGL is combined with ERE.

\acknowledgments

We are grateful for the support provided by Kiso observatory.  
We wish to thank for T.~D. Brandt for kindly providing their data in digital form as well as  very useful comments and  discussion. 
We also wish to thank for support and useful discussion for T.Yamamoto, T. Kozasa, and  K.Dobashi.
We appreciate the comments and criticisms given by the anonymous referee, which were extremely useful and constructive.
NI is supported by Grant-in-Aid for Japan Society for 
the Promotion of Science (JSPS) Fellows. 
This work has been supported in part by Grants-in-Aid for 
Specially Promoted Research on Innovative Areas (22111503, 24111705), Specially Promoted Research(20001003)
and Scientific Research (22684005) from JSPS.

\appendix

\section{Radiative transfer in a dusty slab}
Here we consider a simple model where a plane-parallel dust slab is illuminated from the back by monochromatic starlight at optical wavelength $\lambda$. We measure optical depth $\tau$ inward toward the plane-parallel slab. Light is produced along a sightline by the scattering of the illuminating starlight and from scattered light in the slab. Here, we assume complete forward-scattering for simplicity. The contribution of an element with optical depth from $\tau$ to $\tau + d\tau$ to the scattered light and the starlight are given by
\begin{equation}
d I_{\nu,sca}(\tau) = [I_{\nu,star}(\tau) + I_{\nu,sca}(\tau)] \omega d\tau - I_{\nu,sca}(\tau) d\tau
\end{equation}
\begin{equation}
d I_{\nu,star}(\tau) = - I_{\nu,star}(\tau) d\tau
\end{equation}
where $I_{\nu,sca}(\tau)$ and $I_{\nu,star}(\tau)$ are the specific intensity of the scattered light and the starlight, respectively, $\omega$ is the albedo. Then the solution of $I_{\nu,sca}(\tau)$ is given by
\begin{equation}
I_{\nu,sca}(\tau) = I_{\nu,star}(0) \exp[-(1-\omega)\tau] [1-\exp(-\omega\tau)]. \label{eq:scat}
\end{equation}

The far-infrared(IR) luminosity $L(IR)$ comes from thermal emission from dust grains which absorb the starlight and the scattered light. The contribution of an element to $L(IR)$ is given by
\begin{equation}
d L_{\nu,IR}(\tau) \propto [I_{\nu,star}(\tau) + I_{\nu,sca}(\tau)] (1-\omega) d\tau.
\end{equation}
By integrating this equation, we have
\begin{equation}
L_{\nu,IR}(\tau)   \propto I_{\nu,star}(0) \{1 - \exp[-(1-\omega)\tau]\}.
\end{equation}

The conversion of the total IR luminosity to the 100$\micron$ luminosity is $L(100\micron$) = 0.52$L(IR)$, which is applicable for models from 0.5 to 1.5 times the local ISRF intensity \citep{brandt2012}. 
Thus we have  
\begin{equation}
b(\lambda) \propto  \exp[-(1-\omega)\tau] [1-\exp(-\omega\tau)] / \{1 - \exp[-(1-\omega)\tau]\}
\end{equation}
where $b(\lambda)$ is the intensity of the scattered light at wavelength $\lambda$ relative to the intensity of 100$\micron$ emission.

\end{document}